\documentstyle[12pt]{article}
\newcommand{\PR}{Phys. Rev. }
\newcommand{\PRL}{Phys. Rev. Lett. }
\newcommand{\PL}{Phys. Lett. }

\newcommand{\calF}{{\cal F}}
\newcommand{\calG}{{\cal G}}
\newcommand{\calL}{{\cal L}}

\newcommand{\veca}{{\bf a}}

\newcommand{\vecB}{{\bf B}}

\newcommand{\vecE}{{\bf E}}

\newcommand{\cdelta}{c_{\delta}}
\newcommand{\sdelta}{s_{\delta}}
\newcommand{\ctheta}{c_{\theta}}
\newcommand{\stheta}{s_{\theta}}
\newcommand{\cphi}{c_{\phi}}
\newcommand{\sphi}{s_{\phi}}

\begin{document}
\baselineskip=16pt

\pagenumbering{arabic}

\vspace{1.0cm}

\begin{center}
{\Large\sf Effective field theory approach to light propagation in
an external magnetic field}
\\[10pt]
\vspace{.5 cm}

{Xue-Peng Hu, Yi Liao\footnote{liaoy@nankai.edu.cn}}

{\small Department of Physics, Nankai University, Tianjin 300071,
China} %

\vspace{2.0ex}

{\bf Abstract}

\end{center}

The recent PVLAS experiment observed the rotation of polarization
and the ellipticity when a linearly polarized laser beam passes
through a transverse magnetic field. The phenomenon cannot be
explained in the conventional QED. We attempt to accommodate the
result by employing an effective theory for the electromagnetic
field alone. No new particles with a mass of order the laser
frequency or below are assumed. To quartic terms in the field
strength, a parity-violating term appears besides the two ordinary
terms. The rotation of polarization and ellipticity are computed
for parity asymmetric and symmetric experimental set-ups. While
rotation occurs in an ideal asymmetric case and has the same
magnitude as the ellipticity, it disappears in a symmetric set-up
like PVLAS. This would mean that we have to appeal to some
low-mass new particles with nontrivial interactions with photons
to understand the PVLAS result.

\begin{flushleft}


\end{flushleft}

\newpage
\section{Introduction}

As a consequence of quantum effects, the neutral photons can
interact and lead to nonlinear phenomena in vacuum in the presence
of an electromagnetic field \cite{heisenberg,adler70,iacopini79}.
Recently the PVLAS experiment has observed the rotation of
polarization \cite{PVLAS05a} and the ellipticity \cite{PVLAS05b}
when passing a linearly polarized laser light in vacuum through a
transverse magnetic field. Surprisingly, the observed ellipticity
is larger by a factor of $10^4$ than predicted in the conventional
QED, and a rotation of similar magnitude has also been observed
which however is not expected at the leading order of weak field
expansion in QED \cite{adler70,adler06,biswas06}.

There have been some theoretical attempts to understand the PVLAS
results. The best motivated might be the interpretation in terms
of axion-like particles \cite{sikivie83,maiani86,raffelt88} as the
axion also offers a potential solution to the strong CP problem
and serves as one of the leading candidates for dark matter.
Nevertheless, its mass and coupling to two photons are strongly
constrained by astrophysical observations; for a recent review,
see, e.g., Ref. \cite{raffelt06}. For instance, the parameters
preferred by the PVLAS results would be in the region excluded by
the negative results of the CAST which searches for axion-like
particles coming from the Sun \cite{cast05}. A straightforward
interpretation thus seems not possible. Solutions to this dilemma
have been suggested that invoke novel suppression mechanisms for
axion emission in stellar medium \cite{masso05,jain05,jaeckel06}
or its propagation from the stars to the Earth \cite{jain06},
para-photons to induce different effective charges to particles in
vacuum and in stellar plasmas \cite{masso06}, or introduce more
specific structure in the photon-(pseudo-)scalar sector
\cite{mohapatra06}. Or more directly, low-mass milli-charged
particles can also contribute via real production or virtual
processes to the quantities observed by PVLAS without conflicting
with astrophysical observations \cite{gies06,ahlers06}.

Instead of introducing more hypothetical low-mass particles and
endowing more complicated properties with them, we consider it
natural to ask whether it is possible to understand the PVLAS
results within the realm of photon interactions. Our idea is to
work with an effective field theory for photons alone at the
energy scale of order eV as specified by the laser light
frequency, and investigate the feasibility to incorporate the
results on the rotation of polarization and ellipticity. The
effective theory generally contains some free parameters that are
ultimately determined by certain fundamental theory defined at a
higher mass scale. If this is feasible, the next step would be to
single out the fundamental theory that can reproduce the
parameters favored by experiments. If it is not, it would be
inevitable to introduce particles of mass lower than an eV with
some exotic properties if the experiments are to be explained
within particle physics.

The paper is organized as follows. In the next section we write
down the most general Lagrangian for electromagnetic fields to the
lowest non-trivial order that is consistent with gauge and Lorentz
symmetries. It turns out to contain a parity-violating term. The
equation of motion for the laser light traversing a transverse
magnetic field is then set up, and its propagation eigenmodes are
found. The eigenmodes are employed in section 3 to study the
propagation of the linearly polarized light in an applied magnetic
field that is transverse to the propagation direction. The
dichroism (rotation of the polarization plane) and birefringence
(ellipticity) are computed for two types of experimental set-ups,
one for a Gedanken experiment with light propagating forward in
one direction and the other for the PVLAS with light travelling
forth and back in a Fabry-Perot cavity. Our results are summarized
and conclusions are made in the last section.

\section{Equation of motion and propagation eigenmodes}

We assume there are no other particles than the photon with mass
of order eV (the laser light frequency) or lower. The gauge and
Lorentz invariant effective Lagrangian that describes low energy
photon interactions is an expansion in the field tensor
$F_{\mu\nu}$ and partial derivatives. For terms with a given
number of $F_{\mu\nu}$, those with additional partial derivatives
are suppressed relative to those without either by the low
frequency of the laser light or the slow variation of the external
electromagnetic field compared to the energy scale of some
underlying theory. We can thus ignore all terms with additional
derivatives.

Working to quartic terms in the electromagnetic field, we have in
unrationalized Gaussian units with $\hbar=c=1$
\begin{equation}
4\pi\calL=-\calF+2\kappa_1\calF^2+2\kappa_2\calG^2+4\kappa_3\calF\calG,
\label{eq_lagrangian}
\end{equation}
where $\calF=\frac{1}{4}F^{\mu\nu}F_{\mu\nu},
~\calG=\frac{1}{4}\tilde{F}^{\mu\nu}F_{\mu\nu}$ and
$\tilde{F}^{\mu\nu}=\frac{1}{2}\epsilon^{\mu\nu\alpha\beta}F_{\alpha\beta}$
with $\epsilon^{0123}=+1$. The identification with the classical
field strengths in vacuum is,
$F^{0i}=-E^i,~F^{ij}=-\epsilon^{ijk}B^k$ with $\epsilon^{123}=+1$;
then, $\tilde{F}^{0i}=-B^i,~\tilde{F}^{ij}=\epsilon^{ijk}E^k$ and
$\calF=\frac{1}{2}(\vecB^2-\vecE^2),~\calG=\vecE\cdot\vecB$. With
four factors of $F$ and $\tilde{F}$, there are two ways to
contract Lorentz indices; one contains two chains of contraction
as shown and the other has only one chain like
$F_{\mu\rho}F^{\rho\sigma}F_{\sigma\alpha}F^{\alpha\mu}$. But the
latter structures are not independent because they can be reduced
to combinations of the former:
\begin{eqnarray}
F_{\mu\rho}F^{\rho\sigma}F_{\sigma\alpha}F^{\alpha\mu}
&=&8\calF^2+4\calG^2\nonumber\\
\tilde{F}_{\mu\rho}F^{\rho\sigma}F_{\sigma\alpha}F^{\alpha\mu}
&=&-4\calF\calG\\
\tilde{F}_{\mu\rho}\tilde{F}^{\rho\sigma}F_{\sigma\alpha}F^{\alpha\mu}
&=&4\calG^2\nonumber
\end{eqnarray}
It is not necessary to consider structures with more tildes as
they are related to the above by $\calF\to-\calF,~\calG\to-\calG$
and thus do not yield independent ones. The above quartic terms
are the most general ones consistent with gauge and Lorentz
symmetry.

It is not possible to form a term with an odd number of $F$ and
$\tilde{F}$ without additional derivatives. Such a term would
violate the charge conjugation symmetry (C). To construct such a
term we would need at least one chain of contraction involving an
odd number of $F$ and $\tilde{F}$ which however vanishes
identically due to antisymmetry of $F$ and $\tilde{F}$. The
quartic coefficients $\kappa_i$ have dimensions of ${\rm
mass}^{-4}$ and should be computed from some underlying theory. In
particular, the $\kappa_3$ term violates the parity P (and thus
CP) and should be presumably small comparing to
$\kappa_1,~\kappa_2$. We would like to emphasize that our
effective Lagrangian describes physics at the energy scale of
order eV and is not suitable as virtual insertions, for instance,
into the electromagnetic moments of the electron and muon which
involve energy scales of at least the lepton masses. Finally, the
leading order terms in the Heisenberg-Euler Lagrangian for the
conventional QED \cite{gies99} is reproduced by
$\displaystyle\kappa_1=\frac{\alpha^2}{45\pi m^4_e},
~\kappa_2=\frac{7}{4}\frac{\alpha^2}{45\pi m^4_e}$ and
$\kappa_3=0$, where $m_e$ is the electron mass.

The equation of motion is
\begin{equation}
\partial_{\alpha}F^{\alpha\beta}=
4\partial_{\alpha}\left[\kappa_1\calF F^{\alpha\beta}
+\kappa_2\calG\tilde{F}^{\alpha\beta} %
+\kappa_3\left(\calG F^{\alpha\beta}
+\calF\tilde{F}^{\alpha\beta}\right)\right]
\end{equation}
Now decompose the field into a quantum part denoted by lower-case
letters plus a classical part denoted by capital ones. They will
be identified with the laser light and the applied magnetic field
respectively. The equation of motion contains terms that are
zero-th, first and second order in the quantum field. The zero-th
order terms specify the applied external field and are of no
interest here. The second order terms are too small compared to
the first order ones. What remains is thus the linearized equation
of motion for the quantum field:
\begin{eqnarray}
\partial_{\alpha}f^{\alpha\beta}&=&\partial_{\alpha}\left\{
\kappa_1\left[2(fF)F^{\alpha\beta}+(FF)f^{\alpha\beta}\right]
+\kappa_2\left[2(\tilde{F}f)\tilde{F}^{\alpha\beta}
+(\tilde{F}F)\tilde{f}^{\alpha\beta}\right]\right.\nonumber\\
&&\left. ~~~~~~+\kappa_3\left[2(\tilde{F}f)F^{\alpha\beta}
+(\tilde{F}F)f^{\alpha\beta}
+2(Ff)\tilde{F}^{\alpha\beta}+(FF)\tilde{f}^{\alpha\beta}\right]
\right\}
\end{eqnarray}
where $(XY)\equiv X_{\mu\nu}Y^{\mu\nu}$ with
$X,Y=f,F,\tilde{f},\tilde{F}$.

Now we specialize to the case of a static, spatially uniform
external magnetic field $\vecB$. Although PVLAS applies a rotating
constant magnetic field, it has been shown \cite{adler06,biswas06}
that it can be treated as static due to its low frequency. We
choose the gauge $a^{\mu}=(0,\veca)$ with $\nabla\cdot\veca=0$ for
the light potential 4-vector. We assume the plane wave light
propagates in the $z$ axis and $\vecB$ points in the $x$ axis.
Then, the only nontrivial equation is for the components in the
$(x,y)$ plane, $a_1,~a_2$, which depend on $(t,z)$ as a plane
wave. After some algebra, it simplifies to
\begin{eqnarray}
(\partial_t^2-\partial^2_z)a_i &=&\frac{1}{2}\chi_1
(\partial_t^2-\partial^2_z)a_i-\chi_1\delta_{i2}\partial^2_za_2
-\chi_2\delta_{i1}\partial^2_ta_1\nonumber\\
&&+\chi_3\partial_t\partial_z\left(\delta_{i2}a_1
+\delta_{i1}a_2\right)
\end{eqnarray}
where $\chi_i\equiv 4\kappa_i\vecB^2$, or in components,
\begin{eqnarray}
(\partial_t^2-\partial^2_z)\left(\begin{array}{c}a_1\\a_2\end{array}\right)
&=&\left(\begin{array}{cc}
\frac{1}{2}\chi_1(\partial^2_t-\partial^2_z)-\chi_2\partial^2_t
&\chi_3\partial_t\partial_z\\
\chi_3\partial_t\partial_z
&\frac{1}{2}\chi_1(\partial^2_t-\partial^2_z)-\chi_1\partial^2_z
\end{array}\right)
\left(\begin{array}{c}a_1\\a_2\end{array}\right)
\end{eqnarray}

For the laser propagating in the $+z$ direction, we can replace
$a_i\to a_ie^{i(kz-\omega t)}$. For $|\chi_{1,2,3}|\ll 1$ which is
the case here, the modification by $\vecB$ is small; then it is
safe to ignore the $(k^2-\omega^2)$ terms and identify $\omega$
with $k$ on the right hand side of the equation. The error of this
approximation is $O(\kappa_i^2\vecB^4)$ which goes beyond the
precision of our working Lagrangian (\ref{eq_lagrangian}). The
eigenvalue equation simplifies to
\begin{equation}
\left(\begin{array}{cc}%
(\omega^2-k^2)+k^2\chi_2&k^2\chi_3\\
k^2\chi_3&(\omega^2-k^2)+k^2\chi_1
\end{array}\right)
\left(\begin{array}{c}a_1\\a_2\end{array}\right)=0
\end{equation}
The dispersion relations are determined by the vanishing
determinant of the coefficient matrix to be
\begin{eqnarray}
\omega_{\pm}&=&k\sqrt{1-\bar{\chi}\pm\sqrt{(\Delta\chi)^2+\chi_3^2}},
\end{eqnarray}
where
\begin{equation}
\bar{\chi}=\frac{1}{2}(\chi_1+\chi_2),~\Delta\chi=\frac{1}{2}(\chi_1-\chi_2),
\end{equation}
and the eigenvectors are accordingly
\begin{eqnarray}
\left(\begin{array}{c}a_1\\a_2\end{array}\right)_{\pm} &=&
\left(\begin{array}{c} \sin\delta_{\pm}\\
\cos\delta_{\pm}
\end{array}\right),
\end{eqnarray}
where
\begin{eqnarray}
\sin\delta_{\pm}&=&\frac{\chi_3}{\sqrt{\left[\Delta\chi\mp\sqrt{
(\Delta\chi)^2+\chi_3^2}\right]^2+\chi_3^2}},\nonumber\\
\cos\delta_{\pm}&=&\frac{\Delta\chi\mp\sqrt{(\Delta\chi)^2+\chi_3^2}}
{\sqrt{\left[\Delta\chi\mp\sqrt{(\Delta\chi)^2+\chi_3^2}\right]^2+\chi_3^2}}
\end{eqnarray}

Two comments are in order. First, in contrast to the QED case, the
propagation eigenmodes are generally not those parallel and
perpendicular respectively to the applied $\vecB$ field. Second,
for the laser propagating in the $-z$ direction, the solutions are
obtained by flipping the sign of $\chi_3$. This is because the
$\chi_3$ term in the original equation of motion is linear in
$z$-derivative. These features arise from the parity-violating
term in our Lagrangian.

\section{Dichroism and birefringence}

Having found the propagation eigenmodes of light in a transverse
constant magnetic field, we can now follow its evolution from a
given initial state. To see the effects of the $\kappa_3$ term
clearly, we consider two experimental set-ups, one for an ideal
case and the other for the PVLAS.

\subsection{One-direction propagation}

Assuming that the laser light is linearly polarized with an angle
$\theta$ to $\vecB=|\vecB|\hat{x}$ before entering the field and
propagates in the $+z$ direction, the potential vector for the
light is
\begin{equation}
\veca(0,z)=a_0e^{ikz}\left(
\begin{array}{c}\cos\theta\\
\sin\theta\end{array}\right)
\end{equation}
Decomposing the light traversing the field as a sum of the
propagation eigenmodes,
\begin{equation}
\veca(t,z)=c_+e^{i(kz-\omega_+t)}\left(\begin{array}{c}
\sin\delta_+\\\cos\delta_+
\end{array}\right)
+c_-e^{i(kz-\omega_-t)}\left(\begin{array}{c}
\sin\delta_-\\\cos\delta_-\end{array}\right),
\end{equation}
the initial condition determines
\begin{equation}
c_{\pm}=\pm
a_0\frac{\cos(\theta+\delta_{\mp})}{\sin(\delta_+-\delta_-)}
\end{equation}
After some algebra, the light exiting the $\vecB$ field at
$(t,z)=(T,L)$ is found to be
\begin{eqnarray}
\veca(T,L)&=&a_0e^{i(kL-\bar{\omega}T)}\left(
\begin{array}{c}r_1e^{i\phi_1}\\r_2e^{i\phi_2}
\end{array}\right)
\end{eqnarray}
with
\begin{equation}
\begin{array}{rcl}
r_1e^{i\phi_1}&=&\cos(T\Delta\omega)\cos\theta
-i\sin(T\Delta\omega)\cos(\theta+\delta)\\
r_2e^{i\phi_2}&=&\cos(T\Delta\omega)\sin\theta
+i\sin(T\Delta\omega)\sin(\theta+\delta)
\end{array}
\end{equation}
where the following quantities are defined
\begin{equation}
\begin{array}{rcl}
\bar{\omega}&=&\frac{1}{2}(\omega_++\omega_-) \approx
k\left(1-\frac{1}{2}\bar{\chi}\right),\\
\Delta\omega&=&\frac{1}{2}(\omega_+-\omega_-)
\approx\frac{1}{2}k\sqrt{(\Delta\chi)^2+\chi_3^2},\\
\cos\delta&=&\frac{\Delta\chi}{\sqrt{(\Delta\chi)^2+\chi_3^2}},\\
\sin\delta&=&\frac{\chi_3}{\sqrt{(\Delta\chi)^2+\chi_3^2}}
\end{array}
\end{equation}
For the laser entering the $\vecB$ field at $(t,z)=(0,0)$, we can
identify $T\approx L$.

The rotation of the polarization is measured by $(\alpha-\theta)$
with
\begin{equation}
\tan\alpha=\frac{r_2}{r_1}=\sqrt{\frac{\cos^2(T\Delta\omega)\sin^2\theta
+\sin^2(T\Delta\omega)\sin^2(\theta+\delta)}
{\cos^2(T\Delta\omega)\cos^2\theta
+\sin^2(T\Delta\omega)\cos^2(\theta+\delta)}}
\end{equation}
while the ellipticity is measured by $\tan\beta$ with
\begin{equation}
\sin 2\beta=\sin 2\alpha\sin(\phi_2-\phi_1)
\end{equation}
For $T\Delta\omega\ll 1$, we generally have
$\tan\alpha=\tan\theta+O((T\Delta\omega)^2)$; i.e., there is no
rotation to $O(T\Delta\omega)$. Using
$\displaystyle\phi_2-\phi_1\approx
T\Delta\omega\frac{\sin(2\theta+\delta)}{\sin\theta\cos\theta}$,
we obtain
\begin{equation}
\beta\approx T\Delta\omega\sin(2\theta+\delta)
\approx\frac{1}{2}\bar{\omega} T\left(\Delta\chi\sin
2\theta+\chi_3\cos 2\theta\right) %
\label{eq_beta}
\end{equation}
Note that the parity-conserving and parity-violating contributions
depend differently on the angle between the polarization of light
and the $\vecB$ field.

Interesting cases occur for $\sin 2\theta=0$ where the above
manipulations do not apply. These are the ideal cases when the
angle $\theta$ can be measured to precision better than the small
quantity $T\Delta\omega$. At $\theta=0$, we have
$\frac{r_2}{r_1}\approx\frac{1}{2}\bar{\omega}T|\chi_3|,~
\phi_2-\phi_1\approx\frac{\pi}{2}{\rm sign}(\chi_3)
+\frac{1}{2}\bar{\omega}T\Delta\chi$ so that
\begin{equation}
\alpha\approx\frac{1}{2}\bar{\omega}T|\chi_3|, ~~
\beta\approx\alpha~{\rm sign}(\chi_3)
\approx\frac{1}{2}\bar{\omega} T\chi_3
\end{equation}
At $\theta=\frac{1}{2}\pi$, we find similarly
\begin{equation}
\alpha\approx\frac{\pi}{2}-\frac{1}{2}\bar{\omega}T|\chi_3|,~~
\beta\approx -\frac{1}{2}\bar{\omega}T\chi_3
\end{equation}
In these cases, the rotation and ellipticity have the same
magnitude but differ in sign when $\chi_3<0$ and originate
completely from the parity-violating term in Lagrangian.

\subsection{Round-trip propagation}

Since the optical effects discussed here are very small, it is not
practical in the laboratory experiments to carry out a one-way
propagation of the laser light. The PVLAS experiment employs a
high-finesse Fabry-Perot cavity that allows the light to propagate
forth and back about $N\sim 4.4\times 10^4$ times before exiting
the magnetic field and thus prolongs the optical path
significantly. Our previous discussion assumes a linearly
polarized initial state without ellipticity and does not
necessarily apply to the round-trip set-up. This is because the
rotation and ellipticity accumulated from a previous trip serves
as an initial state for the subsequent one and may affect their
further accumulation. Now we treat this circumstance carefully.

Consider the laser light propagating in the $+z$ direction with
the initial condition:
\begin{equation}
\veca(0,z)=a_0e^{ikz}\left(\begin{array}{c}
e^{-i\phi}\cos\theta\\e^{+i\phi}\sin\theta\end{array}\right)
\end{equation}
On exiting the $\vecB$ field, it becomes
\begin{equation}
\veca(T,L)=a_0e^{i(kL-\bar{\omega}T)}
\left(\begin{array}{c}r_1e^{i\phi_1}\\r_2e^{i\phi_2}\end{array}\right) %
\end{equation}
where \begin{equation}
\begin{array}{rcl}
r_1e^{i\phi_1}&=&\left[C\cphi\ctheta-S\sphi c_{\theta-\delta}\right] %
-i\left[S\cphi c_{\theta+\delta}+C\sphi\ctheta\right],\\
r_2e^{i\phi_2}&=&\left[C\cphi\stheta -S\sphi
s_{\theta-\delta}\right] +i\left[S\cphi
s_{\theta+\delta}+C\sphi\stheta\right]
\end{array}
\end{equation}
and
$C=\cos(T\Delta\omega),~S=\sin(T\Delta\omega),~c_{\phi}=\cos\phi,~s_{\phi}=\sin\phi$,
etc. Denoting
$t_{\theta}=\tan\theta,~\rho=\cdelta\tan(T\Delta\omega),~\sigma=\sdelta\tan(T\Delta\omega)$,
we have
\begin{equation}
\frac{r_2e^{i\phi_2}}{r_1e^{i\phi_1}}=
\frac{(1+i\rho)t_{\theta}e^{i2\phi}+i\sigma} {(1-i\rho)+i\sigma
t_{\theta}e^{i2\phi}}
\end{equation}
For each single trip in the Fabry-Perot cavity, $|\rho|,~|\sigma|$
are tiny. Assuming $t_{\theta}\le 1$, we can expand the above in
$\rho,~\sigma$ (shifting to $t^{-1}_{\theta}e^{-i2\phi}$ if
$t_{\theta}>1$):
\begin{equation}
\frac{r_2e^{i\phi_2}}{r_1e^{i\phi_1}}\approx t_{\theta}e^{i2\phi}
\left[1+i2\rho+i2\sigma c_{2\phi}\frac{c_{2\theta}}{s_{2\theta}}
+2\sigma s_{2\phi}\frac{1}{s_{2\theta}} \right]
\end{equation}
which should be identified with
\begin{equation}
t_{\theta+\delta\theta}e^{i2(\phi+\delta\phi)}\approx
t_{\theta}e^{i2\phi}\left[1+i2\delta\phi+\frac{2\delta\theta}{s_{2\theta}}
\right]
\end{equation}
This gives the increments from a single trip in the $+z$
direction:
\begin{equation}
\delta\phi\approx\rho+\sigma c_{2\phi}t^{-1}_{2\theta},~~
\delta\theta\approx\sigma s_{2\phi}
\end{equation}

A trip in the $-z$ direction from the same initial condition will
yield the increments that are obtained by flipping the sign of
$\sigma\propto\sdelta$:
\begin{equation}
\delta\phi\approx\rho-\sigma c_{2\phi}t^{-1}_{2\theta},~~
\delta\theta\approx-\sigma s_{2\phi}
\end{equation}
For a round-trip, the values of $\phi,~\theta$ in the
trigonometric functions multiplying $\sigma$ of the return-trip
increments can be identified with the initial ones because the
pre-factors $\rho,~\sigma$ are already at $O(\chi_j)$. The errors
caused by this are of $O(|\vecB|^4)$ neglected here from the very
start. Then a round trip cancels the $\sigma$ terms and doubles
the $\rho$ term in $\delta\theta$ and $\delta\phi$. There is thus
no rotation from a round trip. The laser light in PVLAS travels in
the cavity an odd number $N$ of times before exit. Since $N$ is
very large, the contribution from the unpaired single trip can be
neglected compared to the $\frac{1}{2}(N-1)\approx\frac{1}{2}N$
times of the round trip. Thus, the final ellipticity is
\begin{equation}
\beta_N\approx\frac{1}{2}N\bar{\omega} L\Delta\chi\sin 2\theta
\end{equation}
where $L$ is the length of a single pass in the $\vecB$ field.
Note that the ellipticity is independent of $\kappa_3$. This
result could also be obtained from eq. (\ref{eq_beta}) if we
remove the $\chi_3$ term that is anti-symmetric in reflection and
multiply by the number of passes $N$. The conventional QED result
is reproduced by substituting the value
$\displaystyle\Delta\chi=-\frac{\vecB^2\alpha^2}{30\pi m^4_e}$:
\begin{equation}
\displaystyle\beta_{\rm QED}\approx
-L\bar{\omega}\frac{\vecB^2\alpha^2}{60\pi m^4_e}\sin 2\theta
\end{equation}

\section{Summary and conclusion}

We have studied the dichroism and birefringence effects of laser
light in a constant magnetic field from the viewpoint of effective
field theory. As we assumed no new particles with a mass of order
the laser frequency or below, the effective theory is for the
electromagnetic field alone. We employ gauge and Lorentz symmetry
to construct the quartic terms in the field tensor in the
effective Lagrangian. Besides the standard structures occurring in
QED, there is a new term that violates parity. The coefficients of
the terms are free parameters and are to be determined by certain
underlying theory.

The optical effects of the quartic terms are then explored for the
linearly polarized laser in a transverse magnetic field. Due to
parity violation by the new term, we distinguish experimental
set-ups according to their properties under space reflection. For
the asymmetric set-up, both parity conserving and violating
interactions contribute to the ellipticity, but their maximum
effects occur for different orientations of the laser polarization
with respect to the magnetic field: the parity-conserving terms
reaches the largest effect when the polarization has an angle of
$\pi/4$ to the field while the parity-violating term has the
largest contribution when the polarization is parallel or
perpendicular to the field. Concerning the rotation of
polarization, the parity-conserving terms never contributes to the
order considered here, while the violating term results in a
rotation of the same magnitude as the ellipticity in the ideal
case which however looks not practical for the laboratory
experiment. For a symmetric set-up such as PVLAS, all
parity-violating effects disappear; in particular, the ellipticity
depends only on the difference of the coefficients parameterizing
the parity-conserving interactions as happens in the conventional
QED. If the PVLAS results are confirmed, it seems that we have to
invoke some low mass particles with nontrivial interactions with
photons.

\vspace{0.5cm}

{\bf Acknowledgements} We would like to thank X.-Q. Li and Y.-Z.
Ren for helpful discussions.

{\it Note added} When this paper was being prepared for
publication, a new preprint \cite{kruglov07} appeared in which the
low energy polarization effects in a magnetic field of a
hypothetical charged spin-$s$ particle were studied. The particle
was assumed to be in the representation $(0,s)+(s,0)$ of Lorentz
group \cite{kruglov01}. That approach corresponds in our notation
to the case in which $\kappa_3=0$ and $\kappa_1,~\kappa_2$ are
given by the one-loop QED result of the particle. The conclusion
reached in that approach is consistent with ours for this special
case.


\end{document}